\documentclass[prb,twocolumn,showpacs,superscriptaddress]{revtex4-1}
\usepackage{amsmath}
\usepackage{amssymb}
\usepackage{graphicx}
\usepackage{longtable}%
\usepackage{dcolumn}
\usepackage{bm}
\usepackage[usenames]{color}
\usepackage{ulem}

\begin{document}
\title{Pressure-induced phase transitions and the
tetragonal high-pressure modification of Fe$_{1.08}$Te\\}
\author{C. Koz}
\affiliation{Max Planck Institute for Chemical Physics of Solids,
N\"othnitzer Stra\ss e 40, 01187 Dresden, Germany}
\author{S. R\"o{\ss}ler}
\affiliation{Max Planck Institute for Chemical Physics of Solids,
N\"othnitzer Stra\ss e 40, 01187 Dresden, Germany}
\author{A. A. Tsirlin}
\affiliation{Max Planck Institute for Chemical Physics of Solids,
N\"othnitzer Stra\ss e 40, 01187 Dresden, Germany}
\author{D. Kasinathan}
\affiliation{Max Planck Institute for Chemical Physics of Solids,
N\"othnitzer Stra\ss e 40, 01187 Dresden, Germany}
\author{C.~B\"orrnert}
\affiliation{Max Planck Institute for Chemical Physics of Solids,
N\"othnitzer Stra\ss e 40, 01187 Dresden, Germany}
\author{M.~Hanfland}
\affiliation{European Synchrotron Radiation Facility, 6 Rue Jules Horowitz, 38043 Grenoble
Cedex 9, France}
\author{H. Rosner}
\affiliation{Max Planck Institute for Chemical Physics of Solids,
N\"othnitzer Stra\ss e 40, 01187 Dresden, Germany}
\author{S. Wirth}
\affiliation{Max Planck Institute for Chemical Physics of Solids,
N\"othnitzer Stra\ss e 40, 01187 Dresden, Germany}
\author{U.~Schwarz}
\email{schwarz@cpfs.mpg.de}
\affiliation{Max Planck Institute for Chemical Physics of Solids,
N\"othnitzer Stra\ss e 40, 01187 Dresden, Germany}
\date{\today}
%
%
%
\begin{abstract}
We report the effects of hydrostatic pressure on the
temperature-induced phase transitions in Fe$_{1.08}$Te in the
pressure range 0--3~GPa using synchrotron powder x-ray diffraction
(XRD). The results reveal a plethora of phase transitions. At
ambient pressure, Fe$_{1.08}$Te undergoes simultaneous first-order
structural symmetry-breaking and magnetic phase transitions,
namely from the paramagnetic tetragonal (P4/$nmm$) to the
antiferromagnetic monoclinic (P2$_1$/$m$) phase. We show that, at
a pressure of 1.33 GPa, the low temperature structure adopts an
orthorhombic symmetry. More importantly, for pressures of 2.29 GPa
and higher, a symmetry-conserving tetragonal-tetragonal phase
transition has been identified from a change in the $c/a$
ratio of the lattice parameters. The succession of different
pressure and temperature-induced structural and magnetic phases
indicates the presence of strong magneto-elastic
coupling effects in this material.
\end{abstract}
\maketitle
\section{Introduction}
The recent discovery of superconductivity in a Fe-based layered
system by Kamihara $et~ al.$\cite{kamihara2008iron} opened up new
avenues for research in the field of high transition-temperature
superconductivity. The parent compounds of the Fe-superconductors
display ubiquitous magnetic and structural phase transitions. In
this context, the situation is similar to the cuprates for
which the exact nature of the intricate interplay between
structure, magnetism and superconductivity still remains elusive
after more than two decades of intense research. Since the
electronic and phononic excitations are extremely sensitive to the
inter-atomic distances, high pressure can efficiently be used as a
clean tuning parameter to systematically influence and, hence,
gain insight into these complex ordering phenomena.  The physical
properties of Fe-pnictides and chalcogenides display strong
pressure dependencies.\cite{chu2009high} In the case of the 1111
and 122 families of compounds, pressure suppresses the magnetic
transition temperature $T_{\rm N}$
(Refs.~\onlinecite{Lorenz2008effect,Kumar2008effect}) and
concomitantly enhances the superconducting transition temperature
$T_{c}$, \cite{Takahashi2008} which suggests an intimate
relationship between the two order parameters. Under pressure, some
1111-compounds (e.~g., CaFeAsF) undergo a transition from the orthorhombic 
to lower symmetry
monoclinic phase, \cite{Mishra2011} in contrast to the transition from
orthorhombic to higher symmetry tetragonal phase found in 122-type compounds. 
\cite{Mittal2011} In undoped
BaFe$_{2}$As$_{2}$ and SrFe$_{2}$As$_{2}$, pressure induces
superconductivity with $T_{c}$ as high as
38~K.\cite{Takahashi2008high,Kotegawa2009abrupt,Igawa2009pressue,
colombier2009complete} Pressure-induced superconductivity in the
case of CaFe$_{2}$As$_{2}$ is
controversial.\cite{Torikachvili2008pressure,Yu2009absence}
However, all 122 systems exhibit a tetragonal collapsed
phase that seems to exclude
superconductivity.\cite{Kreyssig2008pressure,Uhoya2010,Kasinathan2011}

Among the different families of Fe-superconductors, the tetragonal
Fe$_{1+y}$Se with $T_{c}$~=~8~K can be considered as a reference
material owing to its archetypical binary atomic
pattern.\cite{Hsu2008} The structure belongs to the tetragonal
$P4/nmm$ space group and consists of edge-sharing FeSe$_{4}$
tetrahedra, which form layers orthogonal to the $c$-axis. The
subtle interplay of structural and physical properties in
Fe$_{1+y}$Se is obvious from the fact that superconducting
Fe$_{1.01}$Se undergoes a structural transition from the
tetragonal to the orthorhombic phase at 90~K while
non-superconducting Fe$_{1.03}$Se does not.\cite{McQueen2009}
Moreover, Fe$_{1.01}$Se displays the largest pressure coefficient
in the family of Fe-based superconductors, with $T_{c}$ raising up
to 37~K under a pressure of
7--9~GPa.\cite{Mizuguchi2008,Medvedev2009,Margadonna2009}
Eventually, $T_{c}$ drops with further increase in pressure, and
the crystal structure becomes hexagonal above a pressure of 25
GPa.\cite{Medvedev2009} In addition, $T_{c}$ of Fe$_{1+y}$Se can
also be enhanced by Te substitution up to a maximum of $T_{c}=
15$~K for Fe$_{1+y}$Se$_{0.5}$Te$_{0.5}$.\cite{Yeh2008, Fang2008,
roessler2010} The bulk superconductivity disappears for higher Te
substitution and no superconductivity has been found so far in
bulk samples of the end-member Fe$_{1+y}$Te. Instead, Fe$_{1+y}$Te
displays a unique interplay of magnetic and structural transitions
in dependence on the amount of excess Fe, which is
presumably accommodated in interstitial sites.
\cite{PhysRevLett.102.247001, Li2009, Hu2009,
rodriguez2011magnetic, roessler2011, Zaliznyak} The single,
first-order magnetic and structural transition to the monoclinic
$P2_1/m$ space group observed at $T \approx 69$~K in Fe$_{1.06}$Te
systematically decreases in temperature down to 57~K with an
increase in $y$ from 0.06 to 0.11. For $y \geq 0.12$, two distinct
magnetic and structural transitions occur: the magnetic transition
takes place at a {\it higher} temperature than the
structural one.\cite{roessler2011} Further, for $y \geq 0.12$, the
low-temperature structure adopts orthorhombic symmetry, $Pmmn$.
\cite{PhysRevLett.102.247001,rodriguez2011magnetic,roessler2011}
This space group $Pmmn$ is a maximal non-isomorphic subgroup of $P4/nmm$
with index 2. In turn, the space group $P2_1/m$ of the
\begin{figure}[t]
\centering \includegraphics[width=7.0 cm,clip]{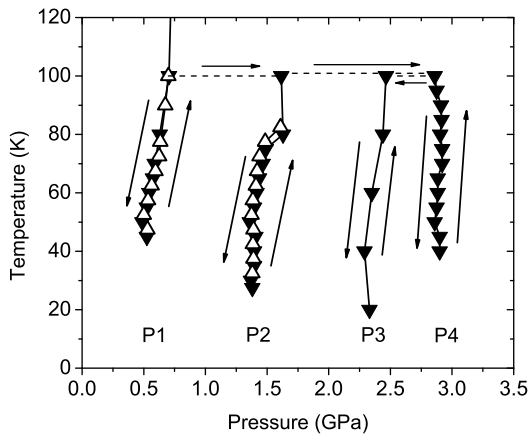}
\caption{Experimental protocol: variation of pressure ($P$)
within the four series (referred to as P1--P4) upon changing
temperature ($T$) during the diffraction experiments.
$\blacktriangledown$ represent $P-$points at which diffraction
data were collected upon cooling, $\triangle$ mark those measured
upon increasing temperature. The temperature-pressure path followed in 
our experiment is indicated by arrows.} \label{fig1}
\end{figure}
\begin{figure}[b!]
\centering \includegraphics[width=6.8 cm,clip]{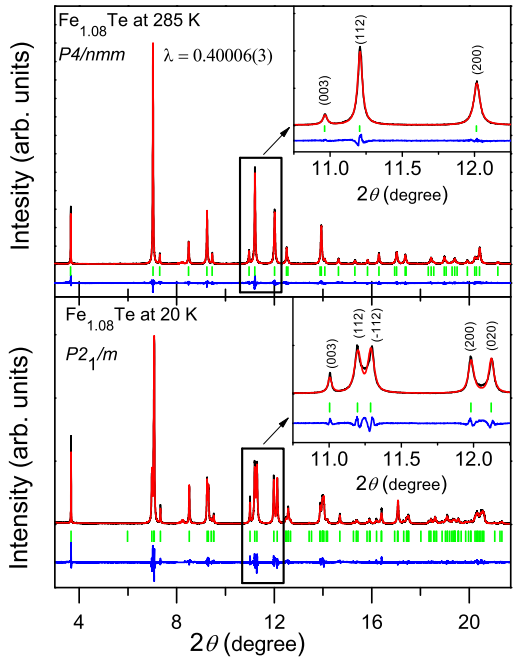}
\caption{Refined synchrotron powder x-ray diffraction patterns of
Fe$_{1.08}$Te at temperatures above (285 K) and below (20 K) the
phase transition ($T_{\rm s} \sim 65$~K) at ambient pressure.}
\label{fig2}
\end{figure}
monoclinic arrangement is a maximal non-isomorphic subgroup of
this orthorhombic variety with index 2. 

So far, high-pressure structural investigations on
Fe$_{1+y}$Te are limited to ambient
temperatures.\cite{Zhang2009,Jorgensen2009} A pressure-induced
tetragonal lattice collapse has been reported for Fe$_{1.05}$Te
and Fe$_{1.087}$Te at 300 K at a pressure of about
4~GPa.\cite{Zhang2009,Jorgensen2009} This collapsed
tetragonal phase was found to be stable up to a pressure of
10~GPa. However, the magnetic and resistive anomalies observed in
a high-pressure study of FeTe$_{0.92}$ (corresponding to
Fe$_{1.086}$Te, {\it cf.} Ref.~\onlinecite{note}) by Okada
{\it et~al.}\cite{okada2009successive} suggested the presence of
two pressure-induced phases at low temperatures. This succession
of phase transitions resembles the result\cite{roessler2011}
obtained at ambient pressure but for higher Fe-content, $y =
0.13$. In order to clearly cross-correlate the influences on the
structure exerted by either pressure or Fe excess, we have chosen
to investigate a
\begin{table}[t]
\caption{Parameters of crystal structures and refinements, atomic
positions and atomic displacement parameters $U_{\text{iso}}$
(in~10$^{-2}$ \r A$^2$) in the tetragonal phase at 285~K and in
the monoclinic phase at 20~K.} 
\label{tab1}
\begin{ruledtabular}
\begin{tabular}{lll}
Temperature (K) & 285 & 20\\
\hline
Space group    & $P4/nmm$       &     $P2_1/m$     \\
 \hspace*{0.28cm}$a$ (\AA)     &     3.82326(4) &     3.83367(8)   \\
 \hspace*{0.28cm}$b$ (\AA)     &    = $a$       &     3.78932(7)   \\
 \hspace*{0.28cm}$c$ (\AA)     &    6.2824(1)   &     6.2594(1)    \\
\hspace*{0.28cm}$\beta$ (degree)  & 90 &  90.661(1)  \\
\hspace*{0.28cm}$R_I/R_P$      &  0.022/0.067   &     0.015/0.095  \\
Number of reflections &  121    &      361 \\
Refined  parameters for &&\\
profile/crystal structure & 20 / 6 & 29 / 11 \\
Atomic parameters& &                                \\
\hspace*{0.28cm}Fe1     &   $2a$ ($\frac34$,$\frac14$,0) & $2e$($x$,$\frac14$,$z$) \\
        &   $U_{\text{iso}}$ = 0.93(2)   &  $x$ = 0.7379(4)           \\
        &                                &  $z$ = $0.0014(3)$        \\
        &                                &  $U_{\text{iso}}$= 0.23(2) \\
\hspace*{0.28cm}Fe2$^a$ & $2c$ ($\frac14$,$\frac14$,$z$) &  $2e$ ($x$,$\frac14$,$z$)  \\
        & $z$ = 0.720(1)                 &  $x$ = 0.238(4)            \\
        & $U_{\text{iso}}$ = 0.9(1)      &  $z$ = 0.719(2)            \\
        &                                &  $U_{\text{iso}}$ = 0.4(2) \\
\hspace*{0.28cm}Te      & $2c$ ($\frac14$,$\frac14$,$z$) &  $2e$ ($x$,$\frac14$,$z$)  \\
        & $z$ = 0.2807(6)               &  $x$ = 0.2431(2)           \\
        & $U_{\text{iso}}$ = 1.08(1)     &  $z$ = 0.2810(1)          \\
        &                                &  $U_{\text{iso}}$ = 0.21(1)
\footnotetext{Atomic displacement parameters, $U_{\text{iso}}$,
and occupancies are intrinsically correlated and, therefore, can
not be refined independently. Rietveld refinements performed with
the nominal composition Fe$_{1.08}$Te  yielded unreasonably small
or even negative values for $U_{\text{iso}}$. Realistic values of
$U_{\text{iso}}$ could be obtained with SOF(Fe2) = 0.09
corresponding to Fe$_{1.09}$Te.}
\end{tabular}
\end{ruledtabular}
\end{table}
sample with $y = 0.08$. For this composition, which is close
to the one used in Ref.\ \onlinecite{okada2009successive}, we
determine the structure with increasing pressure $p \le 3$~GPa and
compare the observed structural transformations to the influence
of chemical composition.

\section{Experimental}
Polycrystalline samples were synthesized by solid state reaction
of Fe (Alfa Aesar, 99.995\%) and Te pieces (Chempur, 99.9999\%)
in glassy carbon crucibles covered with lids. Mixtures of
the target composition were placed in the sample containers and sealed
in quartz ampules under vacuum
(10$^{-5}$ mbar). After heating to 973~K with a rate of
100~K/h, the samples were kept at this temperature for 24 h
before increasing the temperature further up to 1173~K.
The dwelling at 1173~K for 12~h was followed by
\begin{figure}[t]
\centering \includegraphics[width=8 cm,clip]{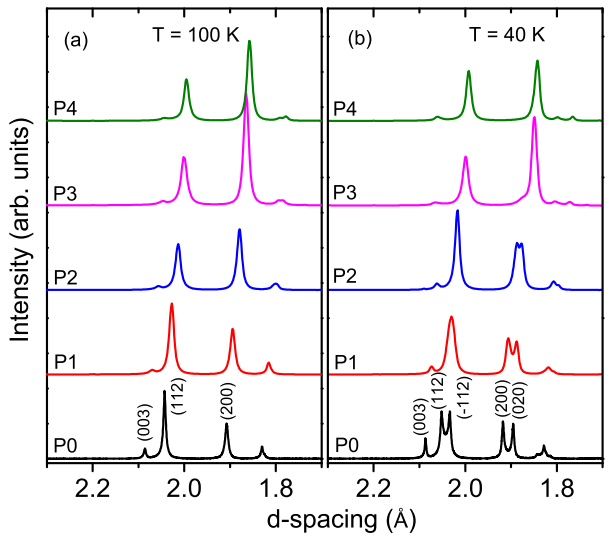}
\caption{Representative high pressure XRD patterns of
Fe$_{1.08}$Te at pressures between ambient (P0) and 2.9~GPa (P4),
(a) at 100~K and (b) at 40~K. } \label{fig3}
\end{figure}
fast cooling to 973~K and annealing for 5~h. Finally, samples were
cooled to room temperature at a rate of 100 K/h. Handling of
starting materials and products was performed in argon-filled
glove boxes. The synthesized samples were characterized by x-ray
powder diffraction using Co~K$\alpha_{1}$ radiation ($\lambda$ =
1.788965 \AA) and wavelength dispersive x-ray (WDX) analysis. The
results clearly show that the samples selected for the
present investigation are single phase with tetragonal symmetry,
$P4/nmm$. According to chemical analysis, the samples contain less
oxygen and carbon than the detection limit of 0.05 mass~\% and
0.06 mass~\%, respectively. As the physical properties of
Fe$_{1+y}$Te depend sensitively on the actual Fe-content $y$,
emphasis was put on its determination. The amount of Fe as
obtained by an inductively-coupled plasma method is systematically
1--2~\% higher than the nominal composition. On the other hand,
WDX analysis reveals an amount of iron that is typically
3--4~\% lower. However, within the estimated experimental error
the results are consistent with the nominal composition. More
importantly and in satisfactory
\begin{figure}[t]
\centering
\includegraphics[width=6.5 cm,clip]{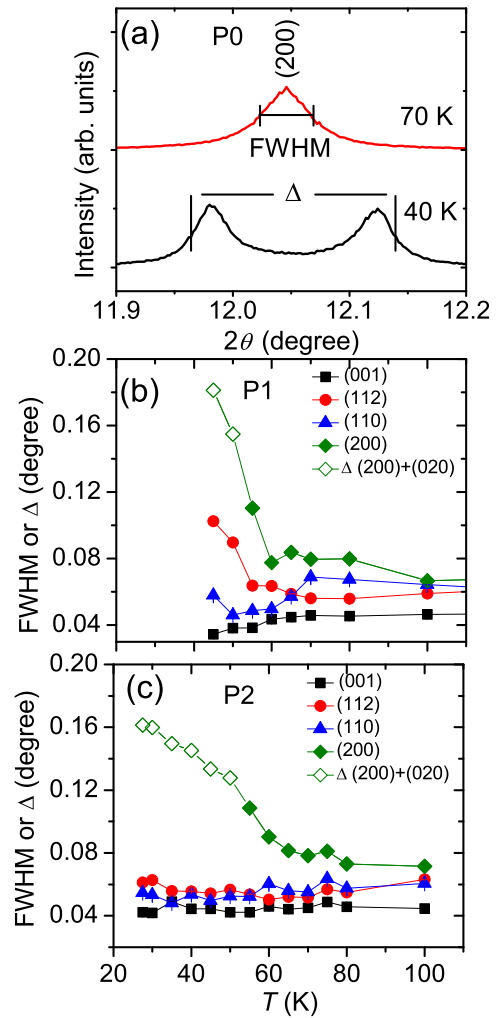}
\caption{Temperature dependence of powder x-ray diffraction peaks
of Fe$_{1.08}$Te. (a) For characterizing the
symmetry-breaking transition, the full width at half maximum
(FWHM), or (for visible splitting) the sum $\Delta$ of separation
of peak maxima plus FWHM, respectively, of selected reflections
are determined. (b) The broadened pattern involving,
e.g., the reflections (112) and (200) evidences mono\-clinic
distortion at low temperatures in the pressure regime P1,
whereas (c) constant values for all peaks except (200)
indicate an orthorombic low-temperature phase at P2. The error bars are smaller than the symbol sizes.}
\label{fig4}
\end{figure}
\begin{figure}[t]
\centering \includegraphics[width=6.9 cm,clip]{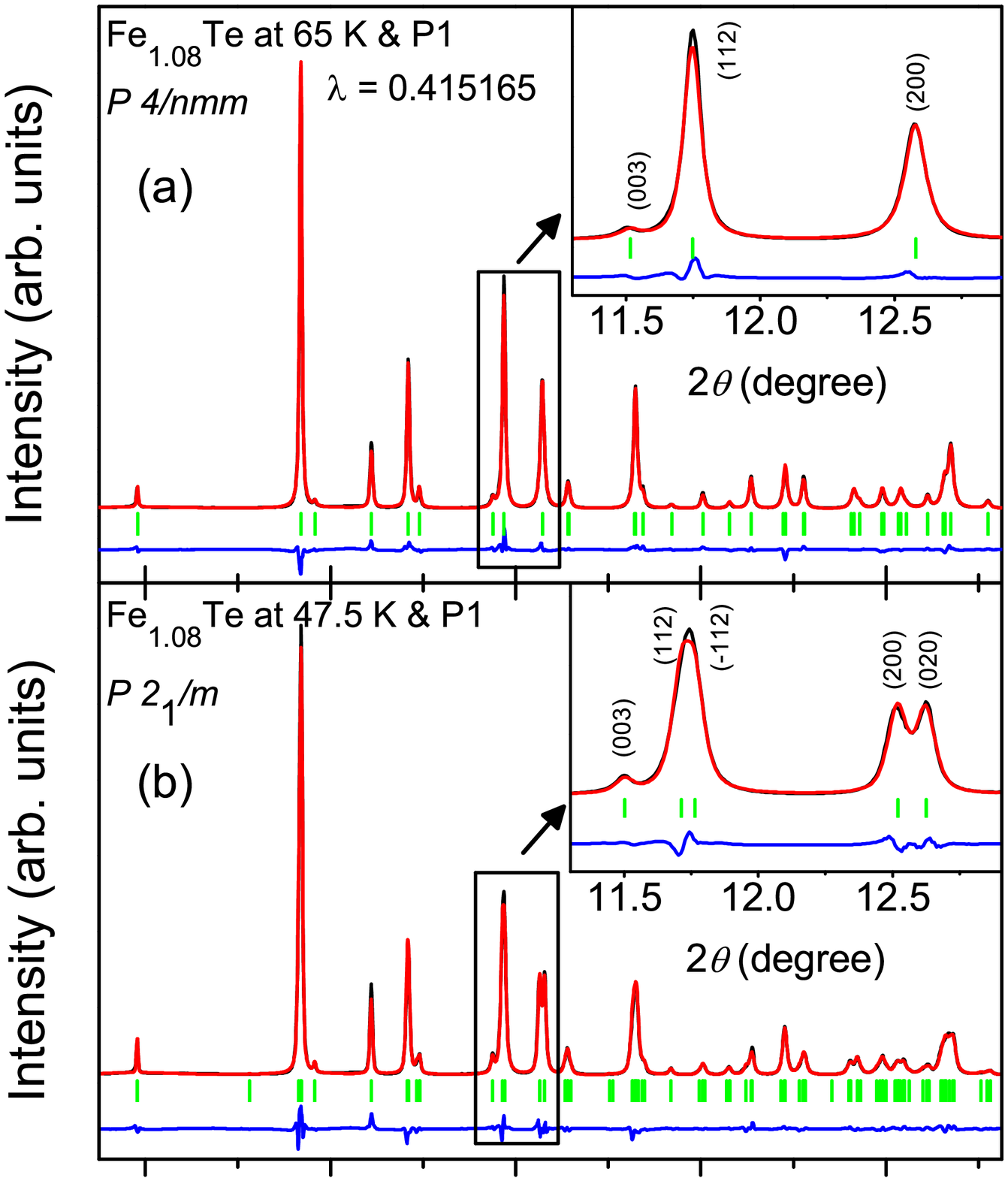}
\centering \includegraphics[width=6.9 cm,clip]{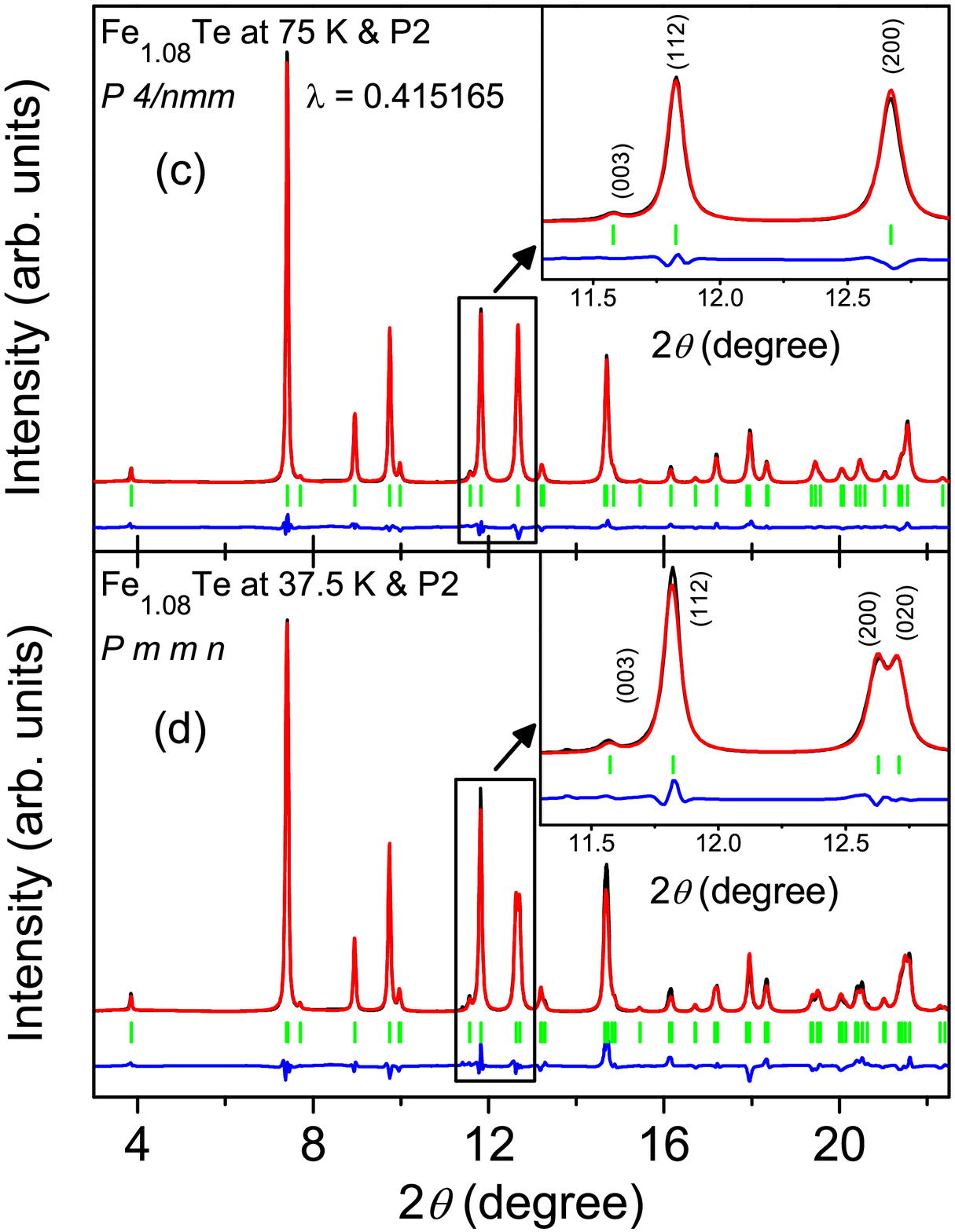}
\caption{Refined synchrotron powder x-ray diffraction patterns of
Fe$_{1.08}$Te for the series P1 at temperatures (a) above
(65 K) and (b) below (47.5 K) the
tetragonal-to-monoclinic transition. (c) and (d) At P2,
a transition from tetragonal to orthorhombic phase is observed.}
\label{fig5}
\end{figure}
\begin{table}[t]
\caption{Parameters of crystal structures and refinements, atomic
positions and atomic displacement parameters $U_{\text{iso}}$
(in~10$^{-2}$ \r A$^2$) at temperatures above and below
the phase transition in the pressure ranges P1 and P2.} 
\label{tab2}
\begin{ruledtabular}
\begin{tabular}{lll}
Temperature (K)                & 65          &    47.5      \\
Pressure P1(GPa)               & 0.58        &    0.53       \\
Space group                    & $P4/ nmm$   & $P2_1/m$\\
\hspace*{0.28cm}$a$ (\AA)                      & 3.7899(1)   &    3.8076(1)  \\
\hspace*{0.28cm}$b$ (\AA)                      & $= a$ &     3.7758(1) \\
\hspace*{0.28cm}$c$ (\AA)                      & 6.2081(2)   &       6.2147(3) \\
\hspace*{0.28cm}$\beta$ (degree) &  90         &     90.354(3) \\
\hspace*{0.28cm}$R_I/R_P$                      &  0.025/0.038&     0.017/0.040  \\
Number of reflections &  39    &      104 \\
Refined  parameters for&&\\
profile/crystal structure & 22 / 6 & 27 / 9 \\
Atomic parameters& &                                \\
 \hspace*{0.28cm}Fe1     &   $2a$ ($\frac34$,$\frac14$,0)  &$2e$($x$,$\frac14$,$z$) \\
         &   $U_{\text{iso}}$ = 0.29(5)    &  $x$ = 0.735(1)        \\
         &                                 &  $z$ = $0.0022(9)$      \\
         &                                 &  $U_{\text{iso}}$= 0.31(6) \\
\hspace*{0.28cm}Fe2$^a$  & $2c$ ($\frac14$,$\frac14$,$z$)  &  $2e$ ($x$,$\frac14$,$z$) \\
         & $z$ = 0.711(4)                &  $x$ = 0.24(1)          \\
         & $U_{\text{iso}}$ = 1.5(5)       &  $z$ = 0.729(4)         \\
         &                                 &  $U_{\text{iso}}$ = 0.30 (fixed) \\
\hspace*{0.28cm}Te       & $2c$ ($\frac14$,$\frac14$,$z$)  &  $2e$ ($x$,$\frac14$,$z$) \\
         & $z$ = 0.2841(1)                 &  $x$ = 0.2480(6)          \\
         & $U_{\text{iso}}$ = 0.19(3)      &  $z$ = 0.2827 (2)         \\
         &                                 &  $U_{\text{iso}}$ = 0.09(3)\\

\hline

Temperature (K)     & 75                &       37.5    \\
Pressure P2(GPa)    & 1.49              &       1.39    \\
Space group         & $P4/nmm $         &       $Pmmn$   \\
\hspace*{0.28cm}$a$ (\AA)           &   3.7620(2)       &       3.7746(4)\\
\hspace*{0.28cm}$b$ (\AA)           &  $= a$      &     3.7506(4)  \\
\hspace*{0.28cm}$c$ (\AA)           & 6.1735(4)         &     6.1757(7)  \\
\hspace*{0.28cm}$\beta$ (degree)  &  90   &     90         \\
\hspace*{0.28cm}$R_I/R_P$             & 0.027/0.036     &     0.051/0.059  \\
Number of reflections &  39             &      62 \\
Refined parameters for &&\\
profile / crystal structure & 25 / 6 & 24 / 6 \\
Atomic parameters& &                                \\
 \hspace*{0.28cm}Fe1    & $2a$ ($\frac34$,$\frac14$,0)  &$2b$($\frac34$,$\frac14$,$z$) \\
        &   $U_{\text{iso}}$ = 0.10(5)  &  $z$ = $0.0141(1) $           \\
        &                               &  $U_{\text{iso}}$= 0.28(7)    \\
\hspace*{0.28cm}Fe2$^a$ & $2c$ ($\frac14$,$\frac14$,$z$)& $2a$ ($\frac14$,$\frac14$,$z$)\\
        & $z$ = 0.710(4)                &  $z$ = 0.699(6)               \\
        & $U_{\text{iso}}$ = 0.8(4)     &  $U_{\text{iso}}$ = 1.00 (fixed)   \\
\hspace*{0.28cm}Te      & $2c$ ($\frac14$,$\frac14$,$z$)&  $2a$ ($\frac14$,$\frac14$,$z$)\\
        & $z$ = 0.2867(1)               &  $z$ = 0.2862(2)              \\
        & $U_{\text{iso}}$ = 0.42(3)    &  $U_{\text{iso}}$ = 0.28(4)   \\
\end{tabular}
\end{ruledtabular}
\end{table}
agreement, Rietveld refinements of high resolution synchrotron
powder x-ray diffraction (XRD) data indicate a composition
Fe$_{1.09}$Te for the nominal composition Fe$_{1.08}$Te (see Table
1). Similar subtle variations of the determined amount of Fe result
also from alternative analysis methods as was reported 
independently.\cite{Zaliznyak} 

Our earlier study \cite{roessler2011} on Fe$_{1.08}$Te revealed a
sharp first-order transition at $T_{\rm s} \sim$ 65~K
in the heat capacity $C_{p}(T)$ accompanied by an anomaly in the
temperature dependence of the electrical resistivity $\rho(T)$ and
the magnetic susceptibility $\chi(T)$, corresponding to a
simultaneous magnetic and structural transition. In order to
investigate this transition in detail,  powder XRD experiments
were performed in an angle-dispersive mode at the beam lines ID31
and ID09A of the ESRF (ID31: $\lambda$ = 0.40006(3) \AA~or
0.39993(3) \AA, ID09A: $\lambda$ = 0.415165 \AA). Temperatures
down to $T = 20$~K, both at ambient and elevated pressure, were
realized utilizing special He-flow cryostats adapted to the
requirements of the diffraction set-up environment. The powdered
samples were taken in a thin-wall borosilicate glass capillary for
ambient pressure measurements (ID31). High pressures were
generated by means of the diamond anvil cell technique. The
samples were placed in spark-eroded holes of pre-indented metal
gaskets, together with small ruby spheres for pressure
determination and liquid helium as a force-transmitting medium
(ID09A). The protocol used for the pressure experiment is
presented in Fig.~1. Lattice parameters were determined using the
program package WinCSD,\cite{WINCSD} refinements of the crystal structures 
were performed on the basis of full diffraction profiles with JANA. \cite{JANA}
In these least-squares procedures, the considerable effects of preferred 
orientation caused by the anisotropy of the crystal structure are accounted for 
by the March-Dollase formalism. \cite{March, Dollase}

\section {Results and discussion}
A full-profile refinement of powder XRD data measured at ambient
pressure confirmed a temperature-induced transformation from
tetragonal Fe$_{1.08}$Te (space group $P4/nmm$ at 285~K) into the
monoclinic phase ($P2_{1}/m$ at 20~K) at low temperature
(Fig.~\ref{fig2}). Consistent with earlier results,
\cite{PhysRevLett.102.247001} the phase transition is obvious from
a clear splitting of Bragg peaks like (112) and (200). Refined
structural parameters at 285~K and at 20~K are presented in
Table~\ref{tab1}.

Powder XRD patterns of Fe$_{1.08}$Te in the region of the (112)
and (200) Bragg peaks recorded for four different pressure values
up to 3~GPa are displayed in Fig.~\ref{fig3}. At a temperature of
100~K [Fig.~\ref{fig3}(a)], the diagrams evidence the stability of
the tetragonal phase within
the complete pressure range. The visible peaks shift upon
increasing pressure, indicating a continuous compression. At
around 2.9~GPa, the lattice parameters $a$ and $c$ are 2.6~\% and
2.0~\% smaller than those at ambient pressure, respectively. This
\begin{figure}[t]
\centering
\includegraphics[width=8 cm,clip]{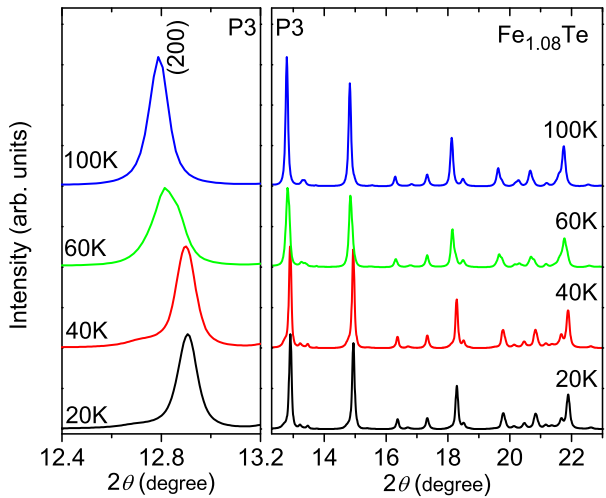}
\caption{Powder x-ray diffraction patterns of Fe$_{1.08}$Te in the
pressure regime P3 at low temperatures. The left part displays the
region of the (200) reflection and the right part an overview of
the broader angular range. The pattern at 100~K shows the
tetragonal high-temperature phase. The shoulder of the (200)
reflection visible at higher angles in the diffraction data taken
at 60~K is assigned to the admixture of a second modification. The
diagrams recorded at 40~K and 20~K exhibit the tetragonal
low-temperature phase. The pronounced shift of, e.g., the (200)
line evidences a significant change of the unit cell parameters
associated with the symmetry-conserving transformation.}
\label{fig6}
\end{figure}
decrease in lattice parameters with pressure at 100~K is slightly
smaller than the recently reported results for
300~K.\cite{Zhang2009,Jorgensen2009}

Upon cooling at only slightly elevated pressures (series P1,
pressure values from 0.31--0.75~GPa dependent on temperature, see
Fig.~1), additional diffraction lines indicate the onset of a
structural change at 55 K. At this temperature, two phases are
identified in the XRD patterns of Fe$_{1.08}$Te. At lower
temperatures, line broadening of the (112) reflection and a
successive splitting of the (200) peak is observed
[Fig.~\ref{fig3}(b)]. In order to characterize the phase
transition, the full width at half maximum (FWHM) as well as
the sum $\Delta$ of separation of peak maxima plus FWHM are
depicted in Fig.~\ref{fig4}. With decreasing temperature,
the refined FWHM value of the (112) peak approximately doubles:
from 0.0512(2)$^\circ$ at 295~K to 0.1025(1)$^\circ$ at 40~K, Fig.
\ref{fig4}(b). Indexing of the reflections at low temperature
requires monoclinic symmetry compatible with the ambient pressure
low-temperature phase, $P2_{1}/m$. Crystal structure refinements
of both high- and low-temperature modification are shown in
Figs.~\ref{fig5} (a) and (b), and the refined parameter values are
listed in Tab.~\ref{tab2}.

At slightly higher pressure (series P2, 1.38--1.65~GPa), the
broadening of the (112) peak at low temperatures is completely
suppressed (FWHM at 100~K: 0.0613(1)$^\circ$; 27.5~K:
0.0631(2)$^\circ$) while the splitting of the (200) and (020)
Bragg peaks remains clearly visible, Fig.~\ref{fig4}(c). Owing to
the modified XRD pattern, the diagrams measured at temperatures of
55 K and below require an orthorhombic lattice for indexing.
\begin{figure}[t]
\centering
\includegraphics[width=6 cm,clip]{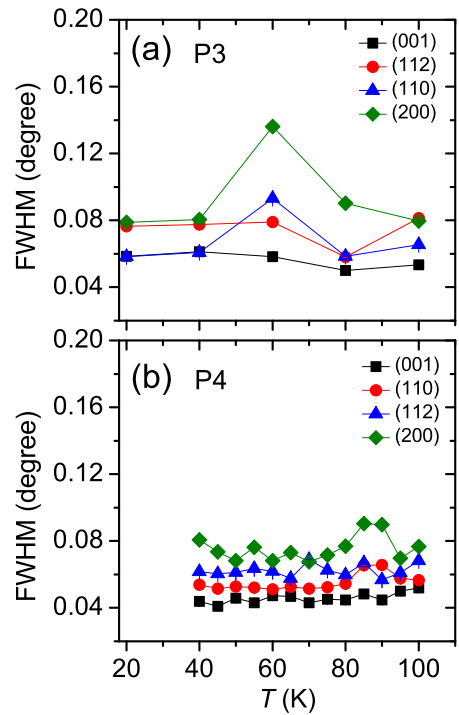}
\caption{Full width at half maximum of selected reflections in the
powder x-ray diffraction diagrams of Fe$_{1.08}$Te as a function
of temperature at a) pressure P3 = 2.29--2.47~GPa and b) pressure
P4 = 2.86--2.92~GPa. The increased values around 60 K in the
pressure range P3 or 90 K for P4 are attributed to transitions
from the tetragonal high-temperature into the tetragonal
low-temperature phase involving two-phase regions. The error bars 
are smaller than the symbol sizes.} 
\label{fig7}
\end{figure}
Systematic extinctions are compatible with space group $Pmmn$.
Consistently, a first Le Bail refinement yields similarly low
values of the residuals as the fit of a monoclinic model. However,
the orthorhombic pattern involves a smaller number of free
parameters and thus, the higher-symmetry $Pmmn$ model is
selected for the crystal structure refinements using full
diffraction profiles. The results for the low- and
high-temperature modifications are visualized in Figs.\
\ref{fig5}(c) and (d), refined parameter values are included
in Tab. \ref{tab2}.
%
\begin{figure}[t]
\centering \includegraphics[width=6.8 cm,clip]{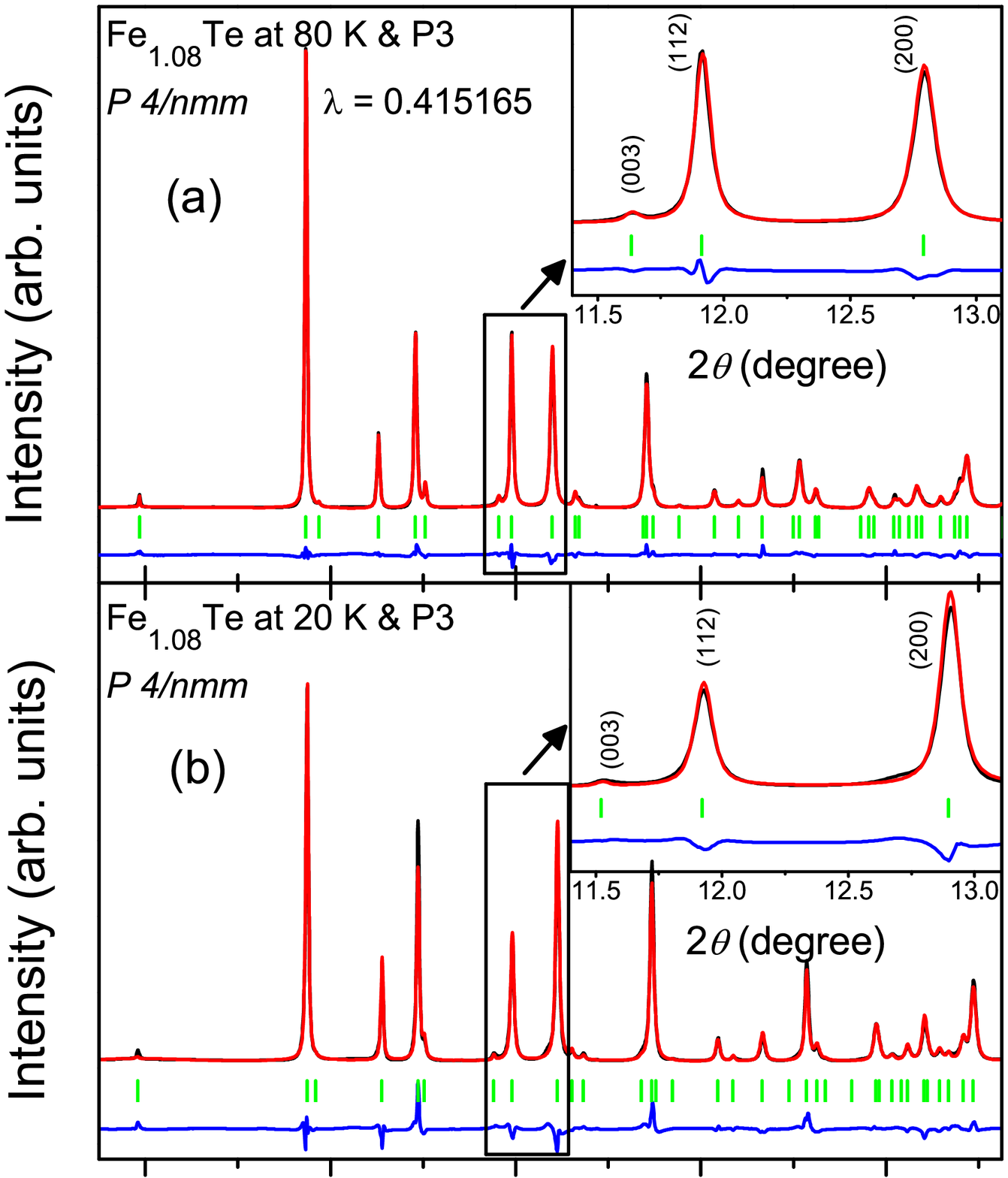}
\centering \includegraphics[width=6.8 cm,clip]{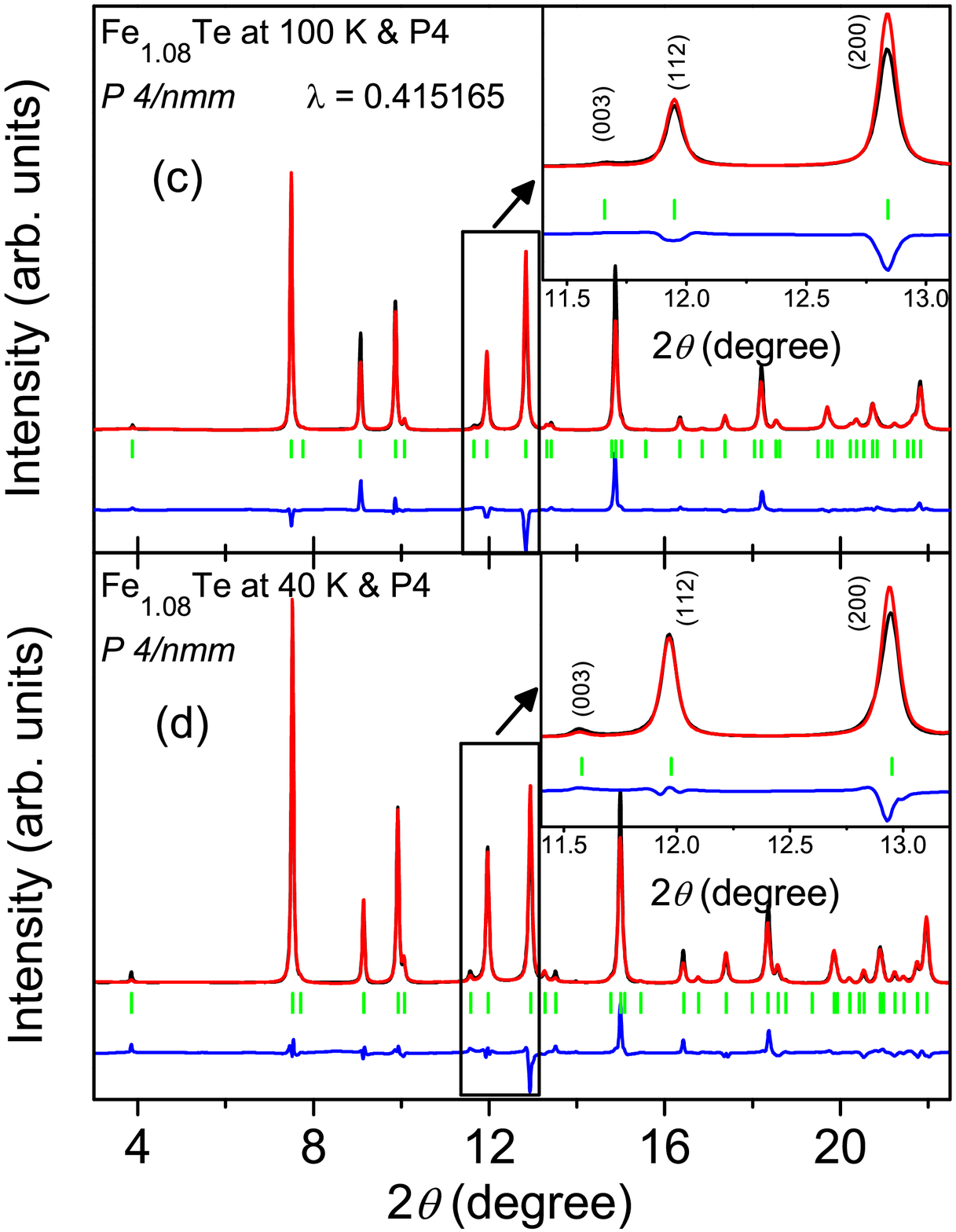}
\caption{Refined synchrotron powder x-ray diffraction patterns of
Fe$_{1.08}$Te for (a), (b) series P3 and (c), (d) series P4.
Shown are results at characteristic temperatures above $[$(a)
80~K, (c) 100~K$]$ and below $[$(b) 20~K, (d) 40~K$]$ the
symmetry-conserving tetragonal-tetragonal phase transition.}
\label{fig8}
\end{figure}
Upon further increase in pressure (series P3, 2.29--2.47~GPa, and
P4, 2.86--2.92~GPa), cooling of the samples induces broadening or
the formation of shoulders for some peaks (Fig.~6). For instance,
in the series P3, the determined FWHM of peak (200) corresponds to
0.0797(2)$^\circ$ at 100~K, then it increases to 0.1362(2)$^\circ$
at 60~K, and finally decreases to 0.0784(1)$^\circ$ at 20~K,
Fig.~\ref{fig7}(a). The patterns of the observed changes in the
series P3 and P4 clearly indicate a temperature-induced phase
transition involving a two-phase region in which both
modifications coexist. Phase coexistence is 
evidenced between 60~K and 40~K at P3, and between 90~K and
80~K at P4, Fig.~\ref{fig7}(b). Co-existing phases in a very large 
pressure range have also been reported in the case of pnictide 
compounds at low temperatures. \cite{Mittal2011, Mishra2011} 
\begin{table}[t]
\caption{Parameters of crystal structures and refinements, atomic
positions and atomic displacement parameters $U_{\text{iso}}$
(in~10$^{-2}$ \r A$^2$) for the pressure ranges P3 and P4.}
\label{tab3}
\begin{ruledtabular}
\begin{tabular}{lll}
Temperature (K)     & 80          &       20    \\
Pressure P3(GPa)    &2.44         &       2.33  \\
Space group         & $P4/nmm $   &   $P4/nmm $ \\
\hspace*{0.28cm}$a$ (\AA)           &   3.7265(1) &       3.6946(1)\\
\hspace*{0.28cm}$b$ (\AA)           &  $= a$      &  $= a$   \\
\hspace*{0.28cm}$c$ (\AA)           & 6.1428(3)   &     6.2010(5)  \\
\hspace*{0.28cm}$\beta$ (degree) & 90 &     90       \\
\hspace*{0.28cm}$R_I/R_P$           &  0.024/0.036  &     0.059/0.073  \\
Number of reflections &  35         &      36 \\
Refined parameters for &&\\
profile / crystal structure & 23 / 5 & 26 / 5 \\
Atomic parameters& &                                \\
\hspace*{0.28cm}Fe1  & $2a$ ($\frac34$,$\frac14$,0) & $2a$($\frac34$,$\frac14$,$z$)\\
     & $U_{\text{iso}}$ = 0.53(5)    &    $U_{\text{iso}}$= 0.43(7)   \\
\hspace*{0.28cm}Fe2  & $2c$ ($\frac14$,$\frac14$,$z$)&  $2c$ ($\frac14$,$\frac14$,$z$)\\
     & $z$ = 0.680(4)              &  $z$ = 0.662(6)                \\
     & $U_{\text{iso}}$ = 0.4(4)     &  $U_{\text{iso}}$ = 0.4 (fixed)     \\
\hspace*{0.28cm}Te   & $2c$ ($\frac14$,$\frac14$,$z$)&  $2c$ ($\frac14$,$\frac14$,$z$)\\
     & $z$ = 0.2911(2)               &  $z$ =  0.2955(3)              \\
     & $U_{\text{iso}}$ = 0.36(3)    &  $U_{\text{iso}}$ = 0.11(5)    \\

\hline

Temperature (K)     & 100    &       40    \\
Pressure P4 (GPa)   & 2.86   &   2.90      \\
Space group         & $P4/nmm$ & $P4/nmm $ \\
\hspace*{0.28cm}$a$ (\AA)           &      3.7131(6) &          3.6835(1) \\
\hspace*{0.28cm}$b$ (\AA)           & $= a$    &     $= a$    \\
\hspace*{0.28cm}$c$ (\AA)           & 6.1316(12)     &       6.1769(5)    \\
\hspace*{0.28cm}$\beta$ (degree) & 90  &     90             \\
\hspace*{0.28cm}$R_I/R_P$           &  0.076/0.089   &     0.077/0.074    \\
Number of reflections &  42          &      42 \\
Refined parameters for &&\\
profile / crystal structure & 23 / 3 & 22 / 5 \\
Atomic parameters& &                                \\
\hspace*{0.28cm}Fe1  &   $2a$ ($\frac34$,$\frac14$,0)&  $2a$ ($\frac34$,$\frac14$,$z$)\\
     &   $U_{\text{iso}}$ = 0.3 (fixed)   &    $U_{\text{iso}}$= 0.95(7)   \\
\hspace*{0.28cm}Fe2  & $2c$ ($\frac14$,$\frac14$,$z$)&  $2c$ ($\frac14$,$\frac14$,$z$)\\
     & $z$ = 0.68(1)              &  $z$ = 0.661(6)                \\
     & $U_{\text{iso}}$ = 0.3 (fixed)     &  $U_{\text{iso}}$ = 0.3 (fixed)     \\
\hspace*{0.28cm}Te   & $2c$ ($\frac14$,$\frac14$,$z$)&  $2c$ ($\frac14$,$\frac14$,$z$)\\
     & $z$ = 0.2922(4)               &  $z$ =  0.2948(3)              \\
     & $U_{\text{iso}}$ = 0.3 (fixed)     &  $U_{\text{iso}}$ = 0.21(4)    \\
\end{tabular}
\end{ruledtabular}
\end{table}
 
A detailed analysis of the line positions revealed that upon 
cooling Bragg peaks like (200)
exhibit strong shifts towards higher $2\theta$ angles whereas
reflections like (00$l$) are reallocated at lower values of
$2\theta$ (see, {\it e.g.}, the (003) peak in the insets to
Figs.~\ref{fig8}). This finding implies that the phase transition
into the low-temperature modification is associated with a
pronounced increase in the ratio $c/a'$ (see below). Comparison of
the diffraction patterns measured at temperatures above and below
this phase transition reveals a close similarity of the
diffraction intensities. Specifically, no evidence for extra
reflections which would indicate, {\it e.g.}, a doubling of
an translation period is observed. Moreover, the diffraction diagrams
of the low-temperature phase can still be indexed assuming
tetragonal symmetry, and the same systematic absences of
reflections are observed for the high- and the
low-temperature phase. The corresponding diffraction symbol is
compatible only with the centrosymmetric space groups $P4/n$ and
$P4/nmm$. Inspection of the occupied Wyckoff positions ($2a$ and
$2c$ in both space groups) immediately reveals that the coordinate
triplets are the same for both choices. Thus, the higher Laue
class was selected for the subsequent refinements. The
least-squares fit results of the structure models to the
diffraction
\begin{figure}[t]
\centering \includegraphics[width=8.5 cm,clip]{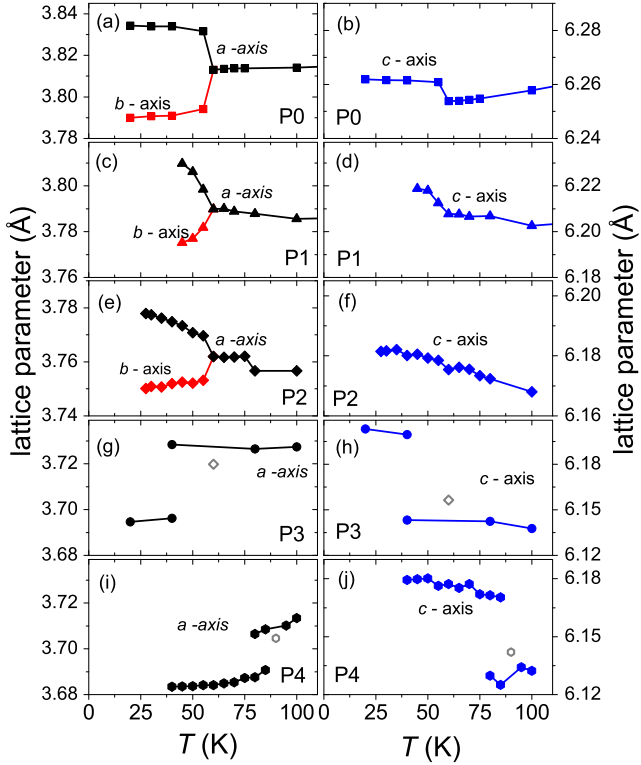}
\caption{Temperature dependence of lattice parameters $a$, $b$ and
$c$ at various pressures up to 3 GPa. (a)--(d) A transition from
tetragonal to monoclinic symmetry is seen at ambient pressure P0
and for pressure P1. (e), (f) For P2, an orthorhombic phase is
found at $T \lesssim 60$~K. Another transition appears to occur at
$T \approx 75$~ K. (g)--(j) A pronounced lattice change within the
tetragonal symmetry is observed for P3 and P4. Open symbols show
average values for mixtures of the high- and low-temperature
phases in P3 and P4. Note a slight temperature-induced decrease of
pressure in the experimental set-up upon cooling, see Fig.~1 and appendix.}
\label{fig9}
\end{figure}
profiles measured above and below the transition at P3 and
P4 are shown in Figs.\ \ref{fig8}(a)--(d); the refined
parameter values are summarized in Tab. \ref{tab3}.

For a comparison of the metrical changes, the temperature
dependence of the lattice parameters obtained from the refinements
at ambient as well as at elevated pressures are summarized in
Figs.~\ref{fig9}(a)--(j). It can be seen that the
symmetry-breaking transitions (tetragonal to monoclinic or
to orthorhombic) are associated with a significantly
anisotropic change of the unit cell dimensions, see
Figs.~\ref{fig9}(a)--(f). In the case of the
symmetry-conserving transition (tetragonal to
tetragonal) the lattice parameter $a$ contracts by
$\approx 1$\% while $c$
\begin{figure}[t]
\centering \includegraphics[width=8 cm,clip]{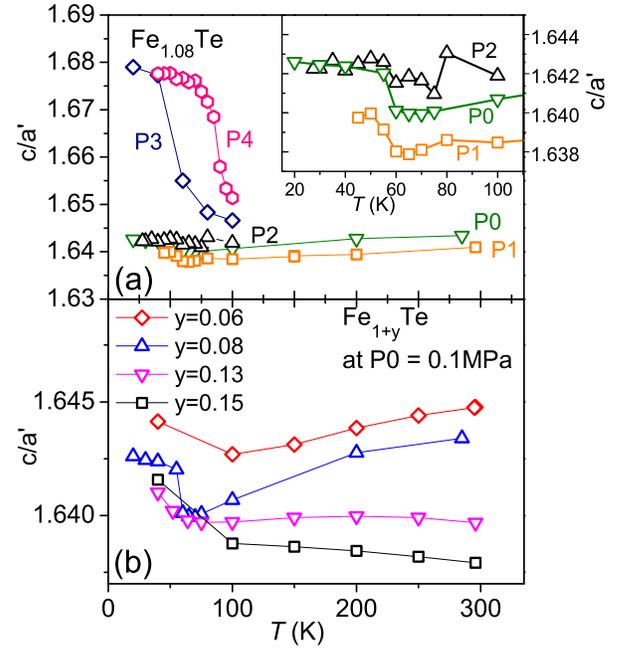}
\caption{Temperature dependence of $c/a'$ where $a' = a$ for
tetragonal symmetry, or equivalently $a' = \frac{1}{2}(a+b)$ for
orthorhombic and monoclinic symmetries at (a) various pressures
on Fe$_{1.08}$Te ({\it i.e.} $y = 0.08$); in the inset, data sets P0--P2 
at low temperatures are magnified for clarity,  and (b) various
amounts $y$ of interstitial Fe at ambient pressure.}
\label{fig10}
\end{figure}%
increases by approximately the same amount upon transforming into
the low-temperature phase, see Figs.~\ref{fig9}(g)--(j).

Putting some emphasis on the similarity between pressure and Fe
excess, the temperature-induced changes of $c/a'$ are compared for
both parameters. Analysis of the ratio $c/a'$ (in which $a' = a$
for tetragonal symmetry, and $a' = \frac{1}{2}(a+b)$ for
orthorhombic and monoclinic symmetries) reveals that
the symmetry-breaking transitions at P0--P2 or compositions
Fe$_{1+y}$Te with $y = 0.06-0.15$ cause only minute changes
of the ratio $c/a'$, whereas the symmetry-conserving
transition gives rise to a significantly more pronounced
alteration, Figs.~\ref{fig10}(a) and (b). 
\section{Summary and Conclusions}
\begin{figure}[t]
\centering \includegraphics[width=8 cm,clip]{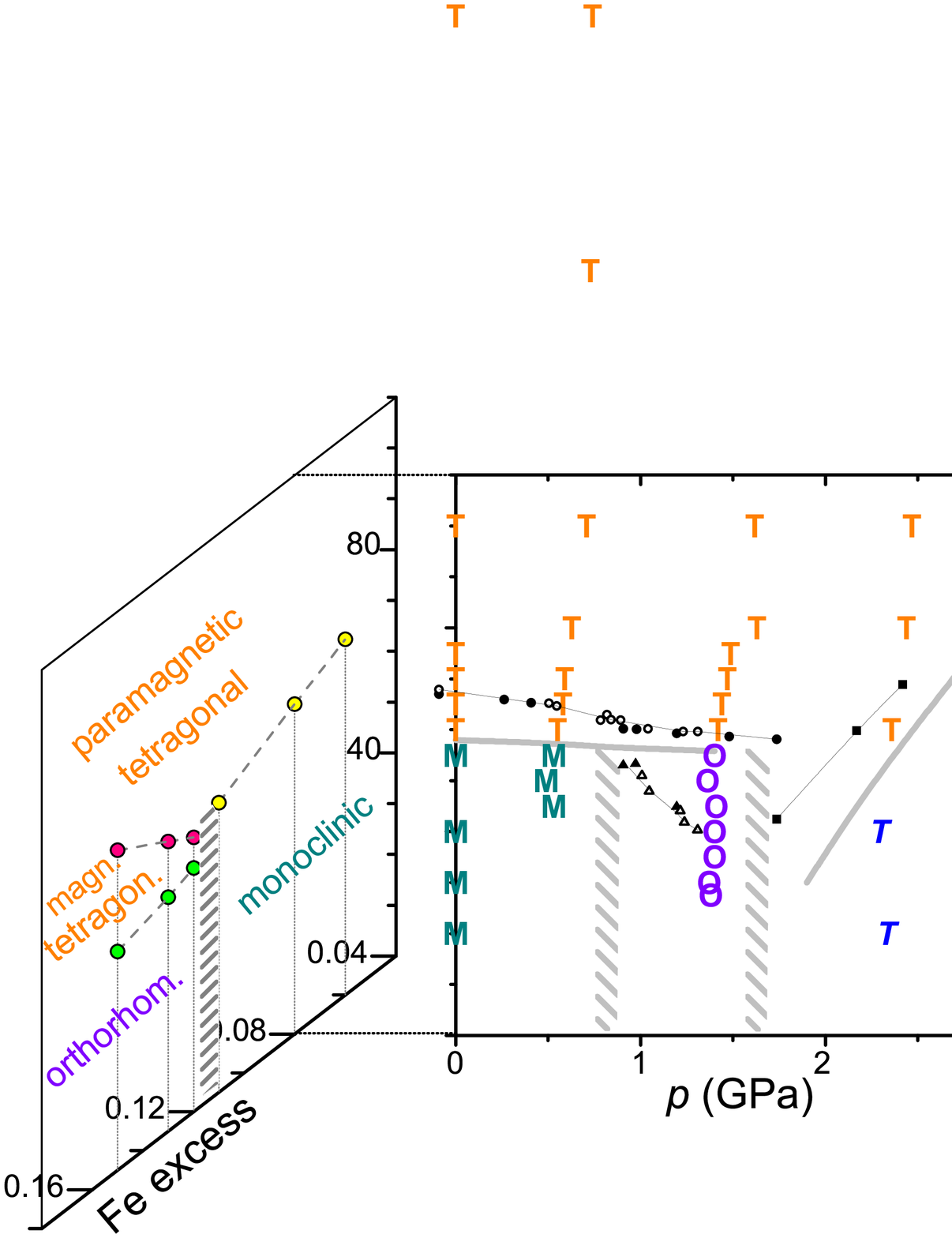}
\caption{Temperature-pressure-composition phase diagram for
Fe$_{1+y}$Te system. Symbols T, M, and O mark temperatures
and pressures of our XRD measurements revealing tetragonal,
orthorhombic and monoclinic phases, respectively. The black
data points indicate anomalies in resistivity, taken from
Ref.~\onlinecite{okada2009successive} for samples Fe$_{1.086}$Te.
Gray regions indicate the existence of structural transitions.}
\label{fig11}
\end{figure}

The anomaly that has been detected \cite{okada2009successive} in
resistivity measurements on Fe$_{1.086}$Te for pressures $p \le
1$~GPa is conjectured to originate from a 
tetragonal--monoclinic phase transition. Our structure investigations
confirm this picture. This phase transition occurs at $T_{\rm s}
\approx 65$~K. At somewhat higher pressures P2 ($\sim
1.4$~GPa), Fig.~\ref{fig9}(e), we clearly resolve a phase
transition into the orthorhombic phase at $T\lesssim 60$~K. 
Yet, the change in $c/a'$ at around 75 K is of similar magnitude 
as the alterations associated to the symmetry-breaking transitions 
at P0 and P1 around 60 K (see inset of Fig.~\ref{fig10} (a)). 
This pressure-driven subtle discontinuity within the tetragonal phase 
is consistent with a change observed for the onset of magnetic 
order in the temperature-composition phase diagram, Fig.~\ref{fig11}.  
At still higher pressures, P3 and P4, we identify another 
symmetry-conserving phase transition. The temperature of this
transition increases with pressure, from $\sim 60$~K at 2.29~GPa
to $\sim 90$~K at 2.9~GPa. There is no indication of the
presence of any orthorhombic or monoclinic phases at these higher
pressures. With this, one might speculate that the unidentified
transition into phase HPII of
Ref.~\onlinecite{okada2009successive} coincides with our
symmetry-conserving phase transition.

Our pressure studies on Fe$_{1.08}$Te here, along with our earlier
investigations\cite{roessler2011} on Fe$_{1+y}$Te samples with
different Fe excess $0.06 \le y \le 0.15$, suggest some analogy
between the influence of pressure and Fe excess. These results
together with results of  Ref.\
\onlinecite{okada2009successive} are summarized in
Fig.~\ref{fig11} for comparison. For small pressures as well as for
small Fe excess $y \le 0.11$ we find a single transition from a
tetragonal into a monoclinic low-temperature phase at roughly
60~K.\cite{note2} At a higher pressure $p \sim 1.5$~GPa or higher
Fe excess $y \ge 0.13$ two successive transitions appear to take
place. Consistently, the transition at lower temperature
($\sim\,$46~K) results in an orthorhombic low-temperature phase.
The second transition at somewhat higher temperature ({\it e.g.}
at 57~K for $y = 0.13$) seems to retain the tetragonal symmetry
but drives the material from a paramagnetic into a magnetically
ordered phase. For even higher pressures $p \gtrsim 2.3$~GPa we
find a symmetry-conserving phase transition. So far, no analogy to
this latter transition has been observed for samples with
increased Fe excess, likely because of the high amount of excess
Fe beyond the homogeneity range of 6--15~\% that would be
required. The exact nature of the magnetism in the teragonal 
high-pressure phase remains to be investigated.

The close similarity of the temperature-composition and the
temperature-pressure phase diagrams suggests a strong
magneto-elastic coupling between the magnetic and structural order
parameters in Fe$_{1+y}$Te. Paul $et~al.$\cite{paul} presented a
mean-field theory, in which symmetry-allowed magneto-elastic
couplings give rise to monoclinic lattice distortion in the
magnetic phase. The magneto-elastic couplings seem to vary with
$y$. For $y~\ge~0.12$, the magnetic structure becomes
incommensurate with respect to the crystal lattice. Neutron
scattering studies report a helical modulation of the magnetic
moments with a temperature-dependent propagation vector.
\cite{PhysRevLett.102.247001, rodriguez2011magnetic} The
structural transition into the orthorhombic phase at lower
temperature takes place only when the magnetic propagation vector
becomes temperature-independent, {\it i.e.}, at the lock-in
transition. \cite{roessler2011} Application of pressure induces
increased overlap of the atomic orbitals
which in turn  tunes the magneto-elastic couplings. This results in 
similar magnetic structures as observed in Fe$_{1+y}$Te
with $y~\ge~0.12$. The microscopic origin of the magnetic and
crystal structures in this regime is not yet theoretically
addressed.

In conclusion, we showed that pressure strongly
influences the phase transitions of Fe$_{1.08}$Te found at low 
temperatures. The temperature-dependent phase transitions can
be successively changed from  low-pressure 
tetragonal--monoclinic to 
tetragonal--orthorhombic followed by tetragonal--tetragonal with
increasing compression. The pressure-dependent phase
transitions closely resemble those induced by excess Fe
composition.

After submission of this article we recognize a report of  an evolution 
of a two step structural phase transition, tetragonal--orthorhombic--monoclinic,
with a two phase (monoclinic + orthorhombic) coexistence at low temperatures 
in Fe$_{1.13}$Te. \cite{Mizuguchi12}
\begin{acknowledgments}
The authors wish to thank Yu.~Grin, K.~Koepernik, U.~K.\
R\"o{\ss}ler and L.~H.\ Tjeng for stimulating discussions. H. R.
thanks SPP 1458 and C. B. thanks CNV--foundation for financial support.
We also acknowledge the ESRF for granting beam time at ID09A and ID31. Experimental 
support of Adrian Hill is highly appreciated.\\
\end{acknowledgments}
\begin{widetext}
\appendix*
\section {Lattice parameters at different temperatures and pressures}
The various phase transitions as outlined in the main text are supported by the results of lattice parameter determinations at different pressures which are summarized in the following Tables IV to VII.
%
\begin{table*}[h!]
\begin{minipage}{17cm}
\renewcommand{\tablename}{TABLE ~$\!\!$}
\caption{\label{table1}Series~P1: Experimental conditions (temperature, pressure) 
and lattice parameters determined by refinement of peak positions using full experimental 
diffraction profiles (LeBail fit). Average differences of temperature 
and pressure before and after the diffraction experiments amount to 0.1(1) K 
and 0.02(1) GPa, respectively.}
\begin{tabular}{l@{\hspace{3em}} c@{\hspace{3em}} c@{\hspace{3em}} c@{\hspace{3em}} c@{\hspace{3em}} c}
\hline
\hline
Temperature (K)	& Pressure (GPa)	& $a$ (\AA)	&	$b$ (\AA)	&		$c$ (\AA)	&	$\beta $(degree)\\
 \hline
		296	&	0.31	&	3.80876(3)	&	-	&6.25002(8) & - \\
		200	&	0.74	&	3.79060(2)	&	-	&6.21435(7)&	-\\
		150	&	0.73	&	3.78497(2)	&	-	&6.20402(7)&	- \\	
	  100	&	0.71	&	3.78560(1)	&	-	&6.20261(7)&	- \\
		80	&	0.63	&	3.78785(2)	&	-	&6.20679(7)&	- \\
		70	&	0.59	&	3.78886(2)	&	-	&6.20654(7)&	- \\
		65	&	0.58	&	3.79004(3)	&	-	&6.2076(1)&	- \\
		60	&	0.55	&	3.78979(2)	&   -	& 6.2078(2)	&-\\
		55	&	0.53	&	3.79845(7)	&   3.78174(7)	& 6.2125(2)	& 90.191(3)\\
		50	&	0.49	&	3.80612(6)	&   3.77699(5)	& 6.2180(2)	& 90.333(2)\\
		45	&	0.53	&	3.80975(7)	&   3.77519(6)	& 6.2187(2)	& 90.399(2)\\
\hline
\hline
\end{tabular}
\end{minipage}
\end{table*}

\begin{table*}[h!]
\begin{minipage}{17cm}
\renewcommand{\tablename}{TABLE ~$\!\!$}
\caption{\label{table1}Series~P2: Experimental conditions (temperature, pressure) and 
lattice parameters determined by refinement of peak positions using full experimental 
diffraction profiles (LeBail fit). Average differences of temperature 
and pressure before and after the diffraction experiments amount to 0.1(1) K 
and 0.02(1) GPa, respectively.}
\begin{tabular}{l@{\hspace{3em}} c@{\hspace{3em}} c@{\hspace{3em}} c@{\hspace{3em}} c@{\hspace{3em}} c}
\hline
\hline
Temperature (K)	& Pressure (GPa)	& $a$ (\AA)	&	$b$ (\AA)	&	$c$ (\AA)	\\
 \hline
		100	&	1.62		&3.75664(1)	&	&	6.1680(1)\\	
		80	&	1.63		&3.75666(1)	&	&	6.17239(9)	\\
		75	&	1.49		&3.76208(2)	&	&	6.17338(8)	\\
		70	&	1.47		&3.76178(1)	&	&	6.17552(8)	\\
		65	&	1.44		&3.76172(2)	&	&	6.17618(8)	\\
		60	&	1.42		&3.76200(2)	&	&	6.17541(9)	\\
		55	&	1.40		&3.76965(6)	&3.75316(7)	&6.1785(2)\\	
		50	&	1.36		&3.77082(6)	&3.75216(6)	&6.1793(2)\\	
		45	&	1.41		&3.77332(5)	&3.75248(5)	&6.1805(2)\\	
		40	&	1.40		&3.77487(5)	&3.75197(5)	&6.1801(2)\\	
  	35	&	1.40		&3.77614(5)	&3.75069(5)	&6.1820(2)\\	
		30	&	1.37		&3.77740(6)	&3.75090(6)	&6.1817(2)\\	
		27.5	&	1.38	&	3.77788(6)&	3.75016(6)&	6.1814(2)	\\
\hline
\hline
\end{tabular}
\end{minipage}
\end{table*}
\begin{table*}[h!]
\begin{minipage}{17cm}
\renewcommand{\tablename}{TABLE ~$\!\!$}
\caption{\label{table1}Series~P3: Experimental conditions (temperature, pressure) and 
lattice parameters determined by refinement of peak positions using full experimental 
diffraction profiles (LeBail fit). Average differences of temperature 
and pressure before and after the diffraction experiments amount to 0.1(1) K 
and 0.02(1) GPa, respectively.}
\begin{tabular}{l@{\hspace{3em}} c@{\hspace{3em}} c@{\hspace{3em}} c@{\hspace{3em}} c@{\hspace{3em}} c}
\hline
\hline
Temperature (K)	& Pressure (GPa)	& $a$ (\AA)	&	$c$ (\AA)	\\
 \hline
		100$^a$	    &	2.47	&	3.72742(2)	& 6.1377(2)\\
		80$^a$	    &	2.44	&	3.72644(2)	& 6.14246(9)\\
		60$^b$	    &	2.36	&	3.71976(2)	& 6.1563(2)\\
		40$^a$	    &	2.29	&	3.7284(5)	& 6.143(2)\\
		40$^c$		  &2.29		&3.69615(2)	  & 6.1996(1)\\
		20$^c$	    &	2.33	&	3.69469(2)	& 6.2033(2)	\\
\hline
\hline
\end{tabular}
\footnotetext{High-temperature phase.}
\footnotetext{Average value for a mixture of the HT- and LT phase since 
             decomposition into the contributions of the components failed.}
\footnotetext{Low-temperature phase.}
\end{minipage}
\end{table*}
\begin{table*}[h!]
\begin{minipage}{17cm}
\renewcommand{\tablename}{TABLE ~$\!\!$}
\caption{\label{table1}Series~P4: Experimental conditions (temperature, pressure) and 
lattice parameters determined by refinement of peak positions using full experimental 
diffraction profiles (LeBail fit). Average differences of temperature 
and pressure before and after the diffraction experiments amount to 0.1(1) K 
and 0.02(1) GPa, respectively.}
\begin{tabular}{l@{\hspace{3em}} c@{\hspace{3em}} c@{\hspace{3em}} c@{\hspace{3em}} c@{\hspace{3em}} c}

\hline
\hline
Temperature (K)	& Pressure (GPa)	& $a$ (\AA)	&	$c$ (\AA)	\\
\hline
		100$^a$	&	2.86		& 3.71343(1)	&6.1323(2)\\	
		95$^a$	&	2.89		& 3.71017(1)	& 6.1341(2)	\\
		90$^b$	&	2.91		& 3.70464(1)	& 6.1421(2)	\\
		85$^a$	&	2.92		& 3.7085(1)	  &6.125(5)	\\
		85$^c$  &	2.92		& 3.69071(6)  & 6.170(4)\\
		80$^a$	&	2.91		& 3.7065(2)	& 6.130(6)	\\
		80$^c$  &	2.91		& 3.68757(3) & 6.171 (2)\\
		75$^c$	&	2.92		& 3.68728(1)	& 6.1719(3)	\\
		70$^c$	&	2.92		& 3.68541(1)	& 6.1773(3)	\\
		65$^c$	&	2.89		& 3.68496(1)	& 6.1753(2)	\\
		60$^c$	&	2.89		& 3.68419(1)	& 6.1773(2)	\\
		55$^c$	&	2.88		& 3.68409(1)	& 6.1764(2)	\\
		50$^c$	&	2.87		& 3.68371(1)	& 6.1802(2)	\\
		45$^c$	&	2.9		&   3.68358(1)	& 6.1798(2)	\\
		40$^c$	&	2.9		& 3.68347(1)	& 6.1794(2)	\\
\hline
\hline
\end{tabular}
\footnotetext{High-temperature phase.}
\footnotetext{Average value for a mixture of the HT- and LT phase since 
             decomposition into the contributions of the components failed.}
\footnotetext{Low-temperature phase.}
\end{minipage}
\end{table*}

\end{widetext}

%
%

%

\begin{thebibliography}{99}
\bibitem{note} Note that some authors use FeTe$_{1-\delta}$ rather
  than Fe$_{1+y}$Te. Since experimental evidence for substantial
  chalcogen vacancies is lacking, the latter notation is more
  frequently used. Here, FeTe$_{0.92}$ corresponds to Fe$_{1.086}$Te.
\end{thebibliography}
\end{document}